\documentclass[amsmath,amssymb,floatfix]{revtex4}

\def\Tr{\textrm{Tr~}}
\def\half{\frac{1}{2}}
\newcommand{\eq}[1]{\begin{equation} #1 \end{equation}}
\newcommand{\eqs}[1]{\begin{eqnarray} #1 \end{eqnarray}}

\def\a{\alpha}

\def\g{\gamma}
\def\d{\delta}
\def\e{\epsilon}

\def\ee{\eta}

\def\l{\lambda}
\def\m{\mu}
\def\n{\nu}
\def\x{\xi}

\def\p{\pi}
\def\r{\rho}
\def\s{\sigma}

\def\ph{\phi}

\def\D{\Delta}

\def\p{\pi}
\def\r{\rho}
\def\s{\sigma}

\def\ph{\phi}

\usepackage[dvips]{graphicx}
\usepackage{amssymb, amsmath, mathrsfs,verbatim, bbm}

\begin{document}
\title{Holographic Renormalisation and the Electroweak Precision Parameters}
\author{Mark Round}
\affiliation{Swansea University,  School of Physical Sciences, \
Singleton Park, Swansea, Wales, UK}
\begin{abstract}We study the effects of holographic renormalisation on an AdS/QCD inspired description of dynamical electroweak symmetry breaking.   Our model is a 5D slice of $\textrm{AdS}_5$ geometry containing a bulk scalar and $SU(2)\times SU(2)$ gauge fields.  The scalar field obtains a VEV which represents a condensate that triggers electroweak symmetry breaking.  Fermion fields are constrained to live on the UV brane and do not propagate in the bulk.  The two-point functions are holographically renormalised through the addition of boundary counterterms.  Measurable quantities are then expressed in terms of well defined physical parameters, free from any spurious dependence on the UV cut-off.  A complete study of the precision parameters is carried out and bounds on physical quantities derived.  The large-$N$ scaling of results is discussed.
\end{abstract}

\maketitle

\section{Introduction}
Collisions at the LHC in CERN are designed to reveal the mechanism of electroweak symmetry breaking (EWSB).  The simplest model of EWSB borrows from the BCS theory of superconductivity but contains a fundamental scalar Higgs boson.  Such models containing a Higgs have long been known to suffer from problems of fine-tuning.  In the case of the standard model Higgs this is the hierarchy problem.  Suggested solutions to the hierarchy problem have often been hindered by the introduction of a new `small hierarchy' problem.  

Should one follow the guide offered by condensed matter more closely, in an attempt to escape fine tuning,  then it would seem preferable to use a fermion condensate to spontaneously break symmetry.  This condensate would be analogous to the Cooper pairs in BCS theory but would break the electroweak symmetry.  

Such a model would then not contain a fundamental Higgs, avoiding its fine-tuning problems.  Early models of electroweak symmetry breaking without a Higgs were based around QCD models of chiral symmetry breaking, and date from the 1970's under the name of technicolor~\cite{TC1,TC2,TC3}.

In its early form technicolor was found to be in violation of electroweak precision tests  and flavour changing neutral current bounds~\cite{PT1, PT2, PT3, PT4}.  To counter such problems the notion of walking technicolor was proposed~\cite{wTC1, wTC3, wTC4, wTC5, wTC6, wTC7, wTC8, wTC9, wTC10,wTC13, wTC14, wTC15, wTC16}.  In addition new computational techniques dealing with the strongly coupled condensate came about with the AdS/CFT correspondence~\cite{AdSCFT1, AdSCFT2, AdSCFT3, AdSCFT4, AdSCFT5} and AdS/QCD~\cite{AdSQCD1, AdSQCD2, AdSQCD3, AdSQCD4, AdSQCD5, AdSQCD6, AdSQCD7, AdSQCD8, AdSQCD9, AdSQCD10}.  A new class of models without a Higgs, so called Higgsless models, which are formulated on a slice of $\textrm{AdS}_5$ were developed~\cite{Higgsless1, Higgsless2, Higgsless3, Higgsless4, Higgsless5, Higgsless6,Higgsless7}.  This gave many new avenues for research into EWSB without a Higgs.

To study a strongly coupled system, like technicolor, one could use AdS/CFT to construct a holographic dual description as is done for example in~\cite{Nunez}.  Alternatively one can adopt a bottom up approach analogous to AdS/QCD, see for example~\cite{Hirn1, Piai}.  It is this later approach we will concentrate on here.  In this approach one begins with a 5D action that has some $SU(2)\times U(1)$ symmetry (rather than the chiral symmetry in AdS/QCD) and a mechanism of inducing symmetry breaking.  The boundary theory of this action is conjectured to be dual to a conformal theory with a condensate that breaks electroweak symmetry.   An example of a physical theory with these properties is the quasi-conformal window of walking technicolor.  Therefore an AdS construction provides a framework with which to study walking technicolor.

Work on technicolor  using an AdS slice has demonstrated models that can satisfy the precision tests unlike the earlier QCD-like technicolor models~\cite{ AdSQCD1, Hirn1, Rattazzi, Grojean, Dietrich, Carone, Hirn2,Weiler}.  Within these new models many papers have discussed how to produce an acceptably small value of the precision parameter $\hat{S}$.  In particular the effect on precision observables of the coupling and anomalous dimension of the techni-condensate have been studied~\cite{Yamawaki}.  Previous work has not discussed the larger set of precision parameters as defined by Barbieri et al.~\cite{LEP} in great detail.  Most previous work concerned field theories that were not renormalised.  The results of such work can contain dependence upon a suprious UV cut-off.   Without counterterms to remove cut-off dependence a study is generally limited to commenting upon Lagrangian parameters and not physical observables (e.g. masses).

The model we will study has already been described elsewhere, see for example~\cite{Piai}.  We aim to combine the ideas of several authors to produce a more complete summary of AdS inspired technicolor models and their predictions for precision parameters.

Our work is organised by initially discussing the model and performing some elementary manipulations, which largely follows previous work [Secs. \ref{Pols} \& \ref{axsol}].  This will  allow us to establish notation and serve as an introduction for those unfamiliar with the ideas of walking technicolor models as studied through the ideas of AdS/CFT.  In Sec. \ref{Divs} we renormalise the 2-point functions through boundary counterterms.  

Using the expressions for the mass spectrum, found in Sec. \ref{D}, a discussion of the large-$N$ counting will be given.  Although the model we study uses a scalar field to induce symmetry breaking, the dual theory it describes is a large-$N$ gauge theory.  We ask how quantities within the $\textrm{AdS}_5$ description we study scale with the increase in $N$ of the technicolor gauge group of the dual theory.   One notable effect of the addition of counterterms is to produce an unexpected set of quantities, each with large-$N$ scaling, compared to large-$N$ QCD intuition.  The quantities we discuss reproduce the expected large-$N$ scaling.  This is the subject of Sec. \ref{N}.  Then we provide expressions for the relevant precision parameters in Sec. \ref{PP}.  Finally in Sec. \ref{Conc} we draw conclusions from our results.

\section{Choice of Model\label{model}}
We begin by giving a brief rationale leading to the action we will study then defining the boundary counterterms.  The main purpose of this section is to define the theory and establish notation.  

Consider a slice of anti-de Sitter (AdS) space with metric,
\eqs{\label{metric}
ds^2 &=& \left(\frac{L}{y} \right)^2 (\ee_{\m\n}dx^\m dx^\n-dy^2),\\
\ee_{\m\n}&=&\textrm{diag}(1,-1,-1,-1).
}and the  $y$ interval defined as $y\in [L_0,L_1]$ .  The energy scale defined by $1/L_0$ is a UV cut-off  and the larger $1/L_1$ an IR cut-off for the theory living on the AdS boundary.

Our purpose is to study electroweak symmetry breaking caused by a condensate.  Therefore into the AdS bulk we place an $SU(2)_L \times SU(2)_R$ gauge theory.   Our notation is that there are two triplets of gauge fields $L^i $ and $R^i$, for $i \in \{1,2,3\}$ indexing a field for each $SU(2)$ generator.  The $SU(2)$ generators will be denoted by $\s^i/2$ with $\s^i$ the Pauli matrices.  The gauge couplings will be denoted by $g_L$($g_R$) for the $SU(2)_L$($SU(2)_R$) fields.  It is useful to define $L_M = L^i_M \s^i/2$ and similar for $R_M$.  A scalar field, $\Phi$, transforming under the bifundamental representation is placed in the bulk.  This scalar obtains a vacuum expectation value (VEV)  so as to produce the symmetry breaking.  Upper case Latin indices will be used for Lorentz indices over the 5D space, for example $M=0,\ldots ,4$.  Using the geometry as defined in eq.(\ref{metric}) the action is,
\eqs{\label{action}
S&=&\iint \sqrt{g}dyd^4x\, g^{MN}\Tr\bigg[ (D_M \Phi )^\dagger D_N \Phi -\half g^{PQ}( L_{MP}L_{NQ}+R_{MP}R_{NQ}) \bigg],\\
D_M \Phi &=& \partial_M \Phi -ig_L L_M \Phi +i g_R \Phi  R_M,\\
L_{MN}&=&\partial_M  L_N- \partial_N L_M - i g_L[L_M,L_N],
}and an analogous field-strength definition for $R_{MN}$.

The field $\Phi$ can be parameterised by two fields.  A scalar $\Sigma$ that obtains a VEV and a set of pseudo-scalars $\ph = \ph^i \s^i/2$,
\eqs{\label{VEV}
\Phi &=& \Sigma e^{i 2\phi / f(y)}\\
\langle \Sigma \rangle &\equiv& f(y) \equiv \frac{\Upsilon}{2}\frac{y^\D}{L_1^\D}\mathbbm{1}
}for $\mathbbm{1}$ a unit $2\times2$ matrix.  $\Sigma$ will not be allowed to fluctuate.  This equation introduces $\Upsilon$, which is a  parameter controlling the strength of the VEV.  In this work $\D$ is a free parameter.  The field $\Phi$ is dual to the techni-quark condensate $\langle \bar{T} T \rangle$ that breaks electroweak symmetry in the spirit of  technicolor models.  The classical scaling of the techni-quark condensate would imply  that $\D=3$.  Yet, such a strongly-coupled condensate will in general have a large anomalous dimension.  We expect that this will reduce $\D$.  In walking technicolor an anomalous dimension of approximately one is required on the basis of phenomeological considerations (see ~\cite{Yamawaki} for a brief overview).   Thus values of $\D$ between two and three are most interesting.    Finally, for our purposes we have chosen not to write the mass term for $\Phi$ in the action.  If a mass term $m^2 \Tr |\Phi |^2$ were added to the action it would be related to $\D$ by,
\eq{
m^2=-\frac{\D (4-\D)}{L^2}.
}

In the 4D boundary theory we study, the leading order expressions for two-point functions are divergent as the UV cut-off is taken to infinity.  In order to send $1/L_0 \rightarrow \infty$ we add to the action eq.(\ref{action}) a term localized on the UV brane,
\eqs{\label{CTaction}
S_{C.T.} &=& -\iint \sqrt{g}dyd^4x\, Z g^{\m\r}g^{\n\s}\bigg( \half \Tr L_{\m\n}L_{\r\s} +\frac{1}{4} R^3_{\m\n}R^3_{\r\s}  \bigg) \d(y-L_0) .
}The metric $g^{\m\n}$ is that given in eq.(\ref{metric}) over the first four spacetime indices, i.e. omitting $y$.  As divergences are encountered $Z$ is adjusted to remedy them.  This is the process of holographic renormalisation~\cite{Skenderis}.  

The chosen counterterm for $R_{MN}$ only refers to $R^3$ thereby explicitly breaking $SU(2)_R$.  Therefore the charged sector of $SU(2)_R$ will not be renormalised.  The boundary theory we study has   towers of resonances with the quantum numbers of a photon, a $Z$ and two $W^\pm$ pairs.  Later it is shown that the chosen form of the counterterms has the effect of reducing the spectrum to that of the standard model for the zeroth modes in the tower (a photon, $Z$ and a $W^\pm$ pair).  In short, the chosen form of the counterterms removes one pair of $W^\pm$ bosons from the lowest mode.  For a detailed discussion of the counterterm see App. \ref{appendix}.

The main computation of this paper will be to find the neutral two-point functions perturbatively.  For such an exercise  it is  better to work with axial and vector fields.  We take linear combinations of $ L$ and $ R $ to define $A $ and $ V$.  The combinations are,
\eqs{\label{combs}
A^M &=&\frac{1}{\sqrt{g_L^2+g_R^2}}(g_L L^M - g_R R^M),\\
V^M&=&\frac{1}{\sqrt{g_L^2+g_R^2}}(g_RL^M+g_LR^M).
}Focusing  on  the vector fields first, the relevant action terms are,
\eq{
S_V = \iint\sqrt{g}dyd^4x \bigg(-\half g^{MP}g^{NQ} \Tr V_{MN}V_{PQ}-\half Z \d(y-L_0)g^{\m\r}g^{\n\s}\Tr V_{\m\n}V_{\r\s}\bigg).
}The definition of $V_{MN}$ is analogous to $L_{MN}$ defined earlier.  However as we will compute the tree-level 2-point functions for the axial and vector fields the tri-linear and quartic couplings will henceforth be neglected, here and in the axial sector.  This is justified by the large-$N$ scaling of the model which is discussed further in Sec. \ref{N}.

The axial terms in eq.(\ref{action}) that we have so far omitted can be found by exchanging $V$ for $A$ in $S_V$ and adding in an additional term originating from $|D_M \Phi |^2$.
\eqs{
S_{A} &=&\iint \sqrt{g}dyd^4x\bigg(-\half g^{MP}g^{NQ} \Tr A_{MN}A_{PQ}-\half Z\d(y-L_0)g^{\m\n}g^{\r\s}\Tr A_{\m\n}A_{\r\s}\nonumber\\
&+& \Tr g^{MN}(D_M\Phi)^\dagger D_N\Phi \bigg)
}

Within $S_V$ and $S_A$ is a mixing between the 4D gauge components $V^\m$($A^\m$) and the $5^\textrm{th}$ component $V^y$($A^y$).  Appropriate gauge fixing terms are added to the action so as to cancel these mixing terms.

The complete action including counterterms and gauge fixing terms is now defined,
\eq{
S_{tot}=S_V+S_A+S_{G.F.}
}with $S_{G.F.}$ the relevant gauge fixing.  The bulk fields $V^M$ and $A^M$ have been decomposed into component fields $(V^\m,V^y)$ and $(A^\m,A^y)$ respectively.  Then a gauge has been chosen so that there are no mixing terms between the two fields $X^\m$ and $X^y$ in both axial and vector cases.  This allows one to consider the two component fields individually which will allow a 4D boundary theory of $A^\m$ and $V^\m$ with energy scale given by $1/y$ to be studied.  Arriving at this theory is the subject of Secs. \ref{Pols} \& \ref{Divs}.

\subsection{Obtaining the 4D Propagators\label{Pols}}
Here we review the procedure of finding 4D propagators in the boundary theory.  In particular we will find the propagators of $V^3$ and $A^3$.  To simply the notation, henceforth $V^3$ and $A^3$ will be abbreviated to $V$ and $A$ respectively.  To find the propagators of the fields, $V^\m$ and $A^\m$ in the boundary theory, one performs the integral in $y$.  This gives the boundary theory with a cut-off $1/L_0$ and a free parameter $L_1$.   Again we work with the vector portion first and apply the standard procedures of quantum field theory.  That is to say, working to quadratic order in the fields we integrate by parts.  It is then possible to read off the boundary conditions such that the field variations vanish on the two walls of the $\textrm{AdS}_5$ space.  The result of this is to enforce,
\eq{
V^y(y) =0,\label{condvy}}for all $y\in[L_0,L_1]$ and \eq{
\partial_y V^\m \vert_{y=L_1}=0.\label{condvm}
}The bulk equation of motion for $V^\m$ can also be found from this procedure. The result is,
\eq{
\left[ \partial^2\ee_{\m\n} - \partial_\m \partial_\n -y\partial_y \frac{1}{y}\partial_y \ee_{\m\n}\right]V^\n=0 .
}

The spectrum is best found by working in Fourier space.   Transforming only in the $x^\m$ dependent  portion of $V^\m$ and extracting the $y$ dependence with a new function $v(q^2,y)$,
\eq{
V^\m(x^\m , y) \rightarrow  v(q^2,y) V^\m (q^2).
}The equation of motion for the transverse components of $V^\m$ can now be re-written as,
\eq{
\left[y\partial_y \frac{1}{y}\partial_y  + q^2 \right]v(q^2,y)=0
}for $q^2$ the square of the 4-momentum of the field $V^\m$.  This equation has a solution consisting of Bessel functions of the first and second kind,
\eq{\label{vecsol}
v(q^2,y) = a_1 y J_1(y \surd q^2) + a_2 y Y_1(y \surd q^2)
}for constants of integration $a_1$ and $a_2$.

If the boundary conditions and equations of motion are substituted back into the action then this leaves a Fourier transformed action for the transverse components of,
\eqs{\label{VBTheory}
S_V+S_{V\, G.F.} &=& \iint dyd^4q\half Z \d(y-L_0) V^\m \frac{L}{y}v^2(q^2,y)q^2\ee_{\m\n}V^\n \nonumber\\
&+&\int d^4q\half\left[ \frac{L}{y}V^\m v(y)\ee_{\m\n}\partial_y v(q^2,y)V^\n\right]_{y=L_0}
}The $y$ integral can now be performed to obtain a 4D theory.   This 4D theory has an action at quadratic order of,
\eqs{
S_{4V} &=&\int d^4q\half V^\m P_{\m\n}\Pi_V  V^\n \\
\Pi_V&=& \mathcal{N}\left[\frac{\partial_y v(q^2,y)}{v(q^2,L_0)} +Z q^2\right]_{y=L_0}
}which defines $\Pi_V$ up to a normalisation $\mathcal{N}$ which we will choose later.

For the gauge fixed axial terms it is simply a case of following through the same methodology as in the vector case.  It will be useful to define,
\eq{
\a^2=\frac{\Upsilon^{2}}{4L_1^{2\D}}(g_L^2+g_R^2)L^2
}Gauge fixing requires that a combination of $A^y$ and $\phi$ vanish, see for example~\cite{DaRold:2005vr} for an explicit expression.  The constraint that,
\eq{\label{bc}
\partial_y A^\m\vert_{y=L_1}=0
}must be enforced as a boundary condition so that the field variations vanish at the boundary.  After Fourier transforming in 4D (which introduces $a(q^2,y)$ analogous to $v(q^2,y)$) the bulk equation of motion for the transverse components is,
\eq{\label{axeqn}
\left(y\partial_y \frac{1}{y}\partial_y +q^2 -\a^2 y^{2\D-2}\right)a(q^2,y)=0.
}If the equation of motion and boundary conditions are substituted back into the gauge fixed action then,
\eqs{\label{ABTheory}
S_A+S_{A\, G.F.} &=& \iint  dy d^4q\half Z\d(y-L_0)A^\m \frac{L}{y}a^2(q^2,y)q^2\ee_{\m\n}A^\n\nonumber\\
&+&\int d^4q\half\left[ \frac{L}{y}A^\m a(q^2,y)\ee_{\m\n}\partial_y a(q^2,y)A^\n\right]_{y=L_0}
}Performing the $y$ integral one obtains the expression for $\Pi_A$,
\eq{
\Pi_A = a^2(q^2,L_0)\frac{L}{L_0}\left[\frac{\partial_y a(q^2,y)}{a(q^2,y)} +Z q^2\right]_{y=L_0}.
}

The manipulations of this section have led to two equations of motion for the profiles $v(q^2,y)$ and $a(q^2,y)$, each with a boundary condition in the IR (eqs. (\ref{condvm}) \& (\ref{bc})).  The propagators, $\Pi_V$ and $\Pi_A$, have been written in terms of the two profiles and a single IR boundary condition fixes the constants of integration from the differential equations.  In the case of the vector profile, an analytical solution of the equation of motion has been given.
\subsection{A Solution of the Axial Equation of Motion\label{axsol}}
Unlike the vector equation of motion, the solution for the axial equation has no closed analytical form.    We now make a digression from the main thrust of our work to address this issue.  

Our approach is to obtain a solution of the propagator rather than for the profile $a(y)$.  To do this, define a new function $P(y)$~\cite{ Yamawaki,Hong} related to the  quantity $\partial_y a(y)/a(y)$ present in $\Pi_A$,
\eq{
P(y)=\frac{1}{y}\partial_y \log a(q^2,y).
}Making $a(y)$ the subject and substituting into eq.(\ref{axeqn}) $P$ satisfies,
\eq{\label{Peqn}
y \partial_y P +y^2 P^2 +q^2=\a^2 y^{2\Delta -2}.
}Now perform an expansion of $P(y)$ in powers of $q^2$,  
\eq{
P(y)=P_0(y)+q^2P_1(y)+q^4P_2(y)+\ldots  
} This gives an infinite series of coupled equations to be solved order by order in $q^2$.  For our purposes only the first three equations will be needed.  They are,
\eqs{
y \partial_y P_0 + y^2 P_0^2 &=&\a^2 y^{2\Delta -2},\\
y \partial_y P_1 +2 y^2 P_0 P_1 &=&-1,\\
y \partial_y P_2 + 2y^2P_0 P_2 + y^2P_1^2&=&0.
}The equation for $P_0$ can be solved for all $\Delta >1$ analytically.  Then the solutions for $P_1$ and $P_2$ can be written out as integrals in terms of the lower order functions.

\eqs{
P_0 (y) &=&\a y^{\D-2}\frac{I_{\frac{1}{\D}-1}\left(\frac{L_1^\D \a}{\D}\right)I_{1-\frac{1}{\D}}\left(\frac{y^\D \a}{\D}\right) - I_{\frac{1}{\D}-1}\left(\frac{y^\D \a}{\D}\right)I_{1-\frac{1}{\D}}\left(\frac{L_1^\D \a}{\D}\right)}{I_{\frac{1}{\D}-1}\left(\frac{L_1^\D \a}{\D}\right)I_{-\frac{1}{\D}}\left(\frac{y^\D \a}{\D}\right) - I_\frac{1}{\D}\left(\frac{y^\D \a}{\D}\right)I_{1-\frac{1}{\D}}\left(\frac{L_1^\D \a}{\D}\right)}\label{P0}
\\
 P_1(y)&=& \exp \left(- 2\int_{1}^y xP_0(x) dx \right)  \int_{y}^{L_1} \frac{\exp \left(2\int_{1}^x z P_0(z)dz \right)}{x}dx\label{P1}\\
 P_2 (y) &=& \exp \left(- 2\int_{1}^y xP_0(x) dx \right) \int_{y}^{L_1} x P_1^2(x)\exp \left(2\int_{1}^x zP_0(z) dz \right) dx
}
The constants of integration have been solved for using the boundary condition in eq.(\ref{bc}).   In the notation of $P$ the boundary condition is $P(L_1)=0$.  This is enforced order by order in $q^2$.  (A technical note; the lower limit of one may seem odd at first but the two occurrences in either $P_1$ or $P_2$ cancel with each other.)

Using the function $P$ the propagator is  written as,
\eqs{
\Pi_A &=& a^2(q^2,L_0)\left[y P(y) +Z q^2\right]_{y=L_0}\\
&=&a^2(q^2,L_0)\left[y (P_0(y)+q^2P_1(y)+q^4P_2(y)+\mathcal{O}(q^6) )  +Z q^2\right]_{y=L_0}.
}The overall normalisation of $a(q^2,L_0)$ will be set in the next section.

The analytical approximation of $\Pi_A$ through an expansion of the function $P(y)$ given above will be used as the basis for the discussion of the divergences and spectrum to follow.  In the case of the precision parameters it is not necessary to go beyond $q^4$ in the expansion of $\Pi_A$.  This is discussed in ~\cite{LEP}.

\subsection{UV Divergences and Counterterms \label{Divs}}
Returning to the expressions derived for $\Pi_V$ and $\Pi_A$, we now consider the divergences as $1/L_0 \rightarrow \infty$.  Such divergences are removed by the counterterm $Z$.

We begin with the simpler vector equations.   The equation of motion has the general solution eq.(\ref{vecsol}) and in the case that $q^2=0$ a constant solution.   A choice must be made on the normalisation of the fields $\mathcal{N}$ to define, for example, the gauge couplings in 4D.  To do this consider the constant solution and in particular its square occurrence, $v^2(q^2,L_0)$, in the 5D gauge fixed action.  Normalise the integral of this term to unity. 
\eqs{
\int_{L_0}^{L_1}\frac{L}{y}v^2(q^2,L_0)(1+Z\d(y-L_0))dy &=& v^2(q^2,L_0)(L\log\frac{L_1}{L_0}+\frac{L}{L_0}Z)
=1\\
\Rightarrow 
\mathcal{N} &=& (L\log\frac{L_1}{L_0}+\frac{L}{L_0}Z)^{-1}\label{norm}.
}To cure the divergence in this expression as $1/L_0 \rightarrow \infty$ take,
\eq{\label{ct}
Z =\frac{L_0}{\e^2}+ L_0\log \frac{L_0}{L_1}
}So that the kinetic terms for the gauge bosons have an $SU(2)_L\times SU(2)_R$ symmetry  we use this normalisation for all the vector and axial modes at the UV boundary (not just the massless photon):   The pre-factors of $v^2(q^2,L_0)$ and $a^2(q^2,L_0)$ in the propagators defined in Sec.~\ref{Pols} now take the value of the massless mode; whose value is defined in eq.(\ref{norm}).  The counterterm $Z$ is the same for all modes.  The choice of using the scale $L_1$ within the definition of $Z$ is equivalent to defining the theory at a renormalisation scale $1/L_1$.

Within the definition of $Z$  a new dimensionless parameter, $\e$, has been introduced.  It is a parameter that measures the relative coupling strength between the zero mode and heavy modes to the UV brane.   As such $\e$ is a physical parameter that we have control over in this model.  The value of $\e$ can be constrained, see Secs. \ref{N} \& \ref{PP}.  By choosing not to renormalise the charged fields of $SU(2)_R$ ($Z=0$)  the corresponding normalisation $\mathcal{N}$ for these zero mode states diverges as $1/L_0 \rightarrow \infty$.  These states are non-normalisable.  This implies that the 4D gauge coupling for these states will go to zero as $1/L_0 \rightarrow \infty$ which decouples the states from the theory.  As a result there are no corresponding light and weakly coupled physical states originating from the $R^{1,2}$ fields.  It is this mechanism that gives rise to the spectrum discussed earlier.

Returning to the general expression for $\Pi_V$ and expanding for large $1/L_0$,
\eq{
\Pi_V =\e^2 \left[ -q^2\left( \g_E +\log \frac{L_0\surd q^2}{2}-\frac{\p}{2}\frac{Y_0(L_1\surd q^2)}{J_0(L_1\surd q^2)}+\mathcal{O}(L_0) \right) +\frac{Z}{L_0}q^2\right]
}where $\g_E$ denotes the Euler-Mascheroni number.  There is a log divergence which is  cured by the counterterm $Z$.  Thus we can safely take  the $1/L_0 \rightarrow \infty$ limit,
\eq{\label{PiV}
\Pi_V = q^2\left[1 -\e^2\left[ \log \frac{L_1\surd q^2}{2} +\g_E - \frac{\p}{2}\frac{Y_0(L_1 \surd q^2)}{J_0 (L_1\surd q^2)} \right]\right]
}This is the renormalised propagator for the vector field.

Knowing the normalisation of $\Pi_V$  the gauge couplings can be extracted.  To do this add into the action eq.(\ref{action}) source terms for the left and right gauge fields localised on the UV brane,
\eqs{
\left(g_L L^3_M J_L^M + g_R R^3_M J_R^M\right)\d(y-L_0).
}Then consider the Fourier transformed 4D currents within the 5D terms written.  These are,
\eqs{
\left(v(q^2,L_0)g_L L^3_\m J_L^\m + v(q^2,L_0)g_R R^3_\m J_R^\m\right)\d(y-L_0)
}This introduces factors of $v(q^2,L_0)$ or equivalently $a(q^2,L_0)$.  The 4D gauge couplings are read off in the left-right basis as,
\eqs{\label{coupling1}
g&=&\frac{\e}{\sqrt{L}}g_L\label{g}\\ \label{coupling2}g'&=&\frac{\e}{\sqrt{L}}g_R\label{g'},
}which defines $g$ and $g'$ in terms of previously introduced parameters.  The coupling $g$ is identified with the $SU(2)$ gauge coupling of the standard model.  Similarly $g'$ is associated to the $U(1)$ of the standard model.  Notice that both these couplings, $g$ and $g'$, are weak and so the boundary theory is perturbative.

In the case of the axial propagator a comment on $\a$ is necessary.  From the definition of $\a$ it is not obvious how $L_0$ dependence may enter.    Therefore we keep $\a$ fixed as $1/L_0 \rightarrow \infty$.  This will be done by adjusting $\Upsilon$ which we now treat as a function of $L_0$ denoted by $\Upsilon(L_0)$.

The small $L_0$ behaviour of $P(L_0)$ to order $q^4$ can be computed and substituted into $\Pi_A$.  Then the limit $1/L_0 \rightarrow \infty$ can be taken and the renormalised axial propagator is found.  The renormalisation of the axial propagator is less straightforward, in comaprison to the vector propagator, due to the use of a series expansion.  We investigate the $P_0$, $P_1$ and $P_2$ expressions for the small $y$ behaviour each in turn.  For simplicity we work in the regime that $L_1^\D \a \ll 1$.  The alternate limit generically leads to a large $\hat{S}$ parameter and is ruled out by experiment.

If one expands $P_0(y)$ for small $y$ the resulting expression at small $L_1^\D \a$ is,
\eqs{\label{P0approx}
P_0(y)&=&\frac{\a^2}{2(\D-1)}( y^{2\D -2} -L_1^{2\D-2} )\nonumber\\
&+& \frac{1}{4}  \frac{\a^4}{\D(2\D-1)}L_1^{4\D -2} +\mathcal{O}[y^a;a>2\D -2, (L_1^\D \a)^6].
}In the limit $1/L_0 \rightarrow \infty$ only the zeroth order term remains so that 
\eq{P_0
(L_0) \rightarrow - \half \frac{L_1^{2\D -2}\a^2}{ \D -1}+\frac{1}{4}  \frac{L_1^{4\D -2}\a^4}{\D(2\D-1)}+\mathcal{O}[ (L_1^\D \a)^6].
} Which is manifestly UV finite.  The higher order terms of $P_0$ in $y$ written in eq.(\ref{P0approx}) will be required to compute the expressions for $P_1$ and $P_2$.  The corrections in $L_1^\D \a$ will be required for a computation of the precision parameters.

If the series expansion in small $y$ of eq. (\ref{P0approx}) is substituted into eq. (\ref{P1}) and the integrals in the definition of $P_1$ performed order by order in $y$ then a series expansion of $y$ is obtained.  To leading order in $y$,
\eqs{
P_1(y) &=& \log \frac{L_1}{y}-\frac{1+\D}{4\D^2}L_1^{2\D}\a^2 +\frac{ (2+\D(2\D^3 -2\D -1))}{16\D (2\D -1)(\D^2 -1)^2} L_1^{4\D} \a^4  +\mathcal{O}(y),
}  The series expansion approach we have adopted will only hold for small $L_1^\D \a$ because only then are the higher order terms in $x$ (as defined in eq.(\ref{P1})) decreasing at the upper bound of the integral.  The counterterm $Z$ renormalises the  expression for $P_1$ as expected.  This series expansion of $P_1$ (including corrections in $y$ not written above) can be substituted into the definition of $P_2$ and again the integrals performed order by order in $y$.  The result is,
\eq{
P_2(L_0) = \frac{1}{4}L_1^2 - \frac{1}{32}\frac{2\D^2-1}{\D^2(\D-1)}L_1^{2\D +2}\a^2
}The leading $L_1^2/4$ term is the same as one would find in the vector case.  The renormalised expression for $\Pi_A$, expanded in $q^2$ and to leading order in $L_1^\D \a$, is
\eq{
\Pi_A(q^2) =\e^2 \left[- \half   \frac{ L_1^{2\D -2} \a^2}{ \D -1} +q^2\left(\frac{1}{\e^2}- \frac{1}{4} \frac{1+\D}{\D^2}L_1^{2\D}\a^2\right) +q^4\frac{1}{4}L_1^2  \right] .
}

The result of this section is that now the propagators are renormalised.  It has been demonstrated that only $Z$ is required as a counterterm.  Extra charged states associated, with $SU(2)_R$ have been removed.  The gauge couplings in 4D have been defined.  Expressions for the axial propagator, as an expansion in $q^2$ have also been found.  This concludes the main computation of our work. 
 
\section{Mass Spectrum \label{D}}

To find the mass spectrum we find the poles of the propagators.  In the vector propagator we expect a massless photon and a tower of excited states which have the same quantum numbers as the standard model photon.  We will limit ourselves to studying the first new state.  As this is the first massive vector state it is referred to as a techni-$\r$.

As expected, $\Pi_V$ vanishes at $q^2=0$ which we identify as the photon.  The higher modes are the roots of,
\eq{
1 -\e^2\left[ \log \frac{L_1\surd q^2}{2} +\g_E - \frac{\p}{2}\frac{Y_0(L_1 \surd q^2)}{J_0 (L_1\surd q^2)}\right] =0
}Suppose the lowest value of $q^2$ that solves this equation is $M_\r^2$, then define $n(\e)$ such that $L_1 M_\r = n(\e)$.   It is instructive to take three special values of $\e$ and solve for $n(\e)$ numerically giving values for $M_\r$ (which is the $\r$ mass),
\eq{\label{rhomass}
M_\r \simeq \left\{\begin{array}{c l}
\frac{2.40}{L_1}&\textrm{for }\e \textrm{ small}\\
\frac{4.15}{L_1}&\textrm{for }\e = 1\\
\frac{4.69}{L_1}&\textrm{for }\e \textrm{ large}
\end{array}\right.
}

The lowest mass solution of the axial propagator is associated with the $Z$ boson.  As there is no closed analytic form for the axial propagator, we assume that the $q^2$ expansion of $P$ is sufficiently convergent  that the first root is well approximated by,
\eq{
m_Z^2=\frac{-P_0(L_0)}{P_1(L_0)+Z/L_0}.
}Therefore the $Z$ mass is approximately,
\eq{\label{Zmass}
m_Z^2 =\frac{2\e^2\a^2 L_1^{2\D -2}}{4(\D-1)-\e^2 \a^2 L_1^{2\D}(1-1/\D^2)}
}

The equations for the charged sector of $SU(2)_{L,R}$  can be found by rotating the unrenormalised propagators, $\Pi_V$ and $\Pi_A$, back into the $L,R$, basis.  Counterterms are applied to the fields corresponding to $L^1$ and $L^2$; so the equations for the light states they produce are the same as those so far derived.   Alternatively by setting $g_R=0$ in the original Lagragian one would be considering an $SU(2)\times U(1)$ theory.  Then the charged equations would be found from the equations so far derived by setting $g_R=g'=0$.  We will make use of the $SU(2)\times U(1)$ theory to see the effect of the custodial symmetry in our model in Sec. \ref{PP}.

\section{Large $N$ Counting\label{N}}
In this section we address the large-$N$ behaviour of quantities in our model.  The action eq. (\ref{action}) is for the specific gauge group of $SU(2)\times SU(2)$ with a scalar field $\Phi$.  The model is conjectured to be dual, or atleast provide a description of, some theory of confining techni-quarks which is at large-$N$.  As $N\rightarrow \infty$ in the techni-quark theory we can ask how the parameters of our model scale with $N$.

To understand the large-$N$ behaviour of our model we attempt to interpret formulae so as to produce a picture consistent with what is known about large-$N$ theories.  The electroweak symmetry breaking VEV in our model is given to the scalar field $\Phi$ by hand.  In a fully descriptive realisation of a confining technicolor theory the VEV of the condensate would arise dynamically and have some large-$N$ scaling.  We can not derive this scaling within our model.  We instead use the scaling of QCD quantites to infer the correct scalings in our model.

For comparative purposes, we briefly recall the $N$ scaling in QCD -- whose dynamics inspire technicolor theories.   In QCD, the pion decay constant, $f_\p^2$, scales with $N$.  The $\r$ mass has a scaling of $N^0$  and its coupling to the pion, $g^2_{\r\p\p}$, scales as $1/N$.  Applied to technicolor, the naive expected scaling would be that the $Z$ and $W^\pm$ masses scale as $N$ and the techni-$\r$ mass scales as $N^0$.

For the model studied here, we use the commonly held scaling of $g_{\r\p\p}$ in QCD to understand how the bulk couplings should scale.  Within the non-bilinear terms of the action eq.(\ref{action}), the coupling between the vector and pion fields  $\pi$ is,
\eq{
g_{\r\p\p}=\frac{1}{\sqrt{L}}\sqrt{g_L^2+g_R^2}\int \sqrt{g}dy v(y)\p(y) \p(y)
}which is proportional to the 5D couplings $g_L$ and $g_R$.   We wish to take $g_{\r\p\p}$ to scale as $1/N$, therefore,
\eq{
\frac{g_L^2}{L}\sim \frac{g_R^2}{L} \sim \frac{1}{N}.
}
These two formulae imply that the bulk gauge couplings go to zero in the large-$N$ limit:  Therefore the assumption of neglecting the tri-linear and quartic vertices when computing the two-point functions in Sec. \ref{Pols} is justified.

In the large-$N$ limit we wish that the 4D gauge couplings $g$ and $g'$ be kept constant.  Therefore eqs. (\ref{g}) \& (\ref{g'}) imply that by taking the  scaling of $\e^2$ to be $N$ we can ensure the 4D gauge couplings remain constant in the large-$N$ limit.

For the particle masses, consider  the limit of $L_1^\D \a$ small as before so that,
\eq{
m_Z^2 = \frac{1}{2}\e^2\frac{ \a^2}{\D-1} L_1^{2\D-2} + \mathcal{O}(\a^4L_1^{4\D})
}After considering the small $\a L_1^\D$ limit we consider $\e^2$ large, corresponding to the large-$N$ limit.  The two limits of $L_1^\D \a$ small and $\e^2$ large do no commute and the alternative order of limits is ruled out by experiment.  From the expectation of QCD we assume that the $Z$ mass scales as $N$.  One would also argue that the condensate scales as $N$.  Within the $Z$ mass  the quantities that scale with $N$ are the VEV $\Upsilon \sim N$ and the gauge couplings $g_{L,R}\sim 1/N$.  These two quantites only occur in the combination of $\a^2$ which must scale as $N^0$ by the given counting.    This means that the $Z$ mass expression scales as $N$ as expected by virtue of the factor of $\e^2$.  Our notation is clarified by noticing that,
\eq{
m_Z^2 \sim \Upsilon^2 (g^2+g'^2)
}by replacing the 5D coupling within $\a^2$ by the 4D ones and so absorbing the factor of $\e^2$.  In this notation it is clear that the scaling and form of $M_Z^2$ is precisely as one would expect;  a VEV multiplied by gauge couplings.

Combining  the $Z$ mass with the $\r$ mass expression,
\eq{
\frac{m_Z^2}{M_\r^2} = \frac{\e^2 \a^2 L_1^2}{ n^2(\e) (\D-1)}
}becomes a ratio that is much less than one and scaling as $N$ in the large-$N$ limit.  The quantities that we have seen the scaling of are,
\eq{
\frac{g_L^2}{L} , \, \frac{g_R^2}{L} \sim \frac{1}{N}, \, \a^2 \sim  1 , \, L_1 \sim 1, \, \e^2 \sim N.
}

It is instructive to see how the large-$N$ scaling enters into the vector propagator $\Pi_V$.  If $\e^2 \sim N$ then $\Pi_V$ can be schematically written out as,
\eq{
\Pi_V = q^2(1-N f(q^2,L_1)),
}for some function $f$.  Originally the Lagrangian of our theory had an overall factor of $N$ which is the factor of $N$ seen infront of $f$.  By introducing the boundary counterterm the kinetic term acquires a factor of $1/N$ from the $1/\e^2$ within the definition of $Z$.  Thus in $\Pi_V$ the $q^2$ term does not scale with $N$.  The $\r$ mass is approximately found by $f(q^2,L_1)=0$ and to leading order has no $N$ dependence.  However at large-$N$ the higher order corrections to the kinetic term, that is the terms in $f$, become dominant.    This picture helps to understand  the scaling of the precision parameters and the role of counterterms as connected to the $N$ scaling in our model.

\section{The Precision Parameters\label{PP}}
Various decay widths, asymmetries and couplings measured to a high precision give rise to a  set of oblique electroweak `precision parameters'~\cite{PT1,LEP}.  Assuming a well-behaved $q^2$-expansion of the propagators in the standard model these parameters  characterise the corrections of higher order operators to the standard model.  The precision parameters we consider are,
\eqs{
\hat{S}&=&\frac{g}{g'}\Pi'_{W^3B}(0) \\
\hat{T}&=&m_W^{-2}\left[\Pi_{W^3W^3}(0)- \Pi_{W^+W^-} (0)\right] \\
\hat{U}&=&\Pi'_{W^+W^-} (0)-\Pi'_{W^3W^3}(0) \\
W&=&\half m_W^2\Pi''_{W^3W^3}(0)\\
X&=&\half m_W^2\Pi''_{W^3B}(0)\\
Y&=&\half m_W^2\Pi''_{BB}(0)
}for $m_W$ the mass of the $W^\pm$ gauge bosons.  Derivatives are with respect to $q^2$ and evaluated at $q^2=0$.   We use the convention that $\Pi_{W^+ W^-}'(0)=\Pi_{BB}'(0)=1$ for all the vectors for $B$ the standard model $U(1)$ and that $\Pi_{W^+ W^-}(0)=m_W^2$.  For a discussion of the experimental bounds of these parameters see~\cite{LEP}.

>From the formulae in Sec. \ref{Divs} an expression for $\hat{S}$ in terms of $L_1$ and $\a$ can be written.   One can  trade away $L_1$ and $\a$ by re-writing them in terms of the $\r$ and $Z$ masses.

Combining these relations the precision parameter $\hat{S}$ is written out as,
\eqs{\label{shat}
\hat{S} &=& \frac{g^2}{g^2+g'^2}  \frac{1}{4}L_1^{2\D} \a^2  \frac{\D +1}{\D^2}   = \half n^2(\e) \frac{\D^2 -1}{\D^2}\frac{m_W^2}{M_\r^2}
}Recall that $M_\r^2 = n^2(\e)/L_1^2$.  The expression for $\hat{S}$ scales with $m_W^2/M_\r^2 \sim N$ in the large $N$ limit.  This computation was also performed in~\cite{AdSQCD1}.  There, the case of $\x \gg1$ ($\x$ as defined in their paper) is discussed.  The definition of $\x$ amounts to a VEV quantity analogous to eq. (\ref{VEV}).  In our work we have chosen to take a combination of the VEV parameter small.  Furthermore we concluded that a large condensate parameter, implied by $\x\gg1$, would lead to large $\hat{S}$.  This is the conclusion reached in~\cite{AdSQCD1} too.  It may also be worth noting that both our work (see later) and ~\cite{AdSQCD1} conclude that dependence upon $\D$ is a weak effect for $\D > 1$.

Taking the experimental upper bound for $\hat{S}$ as $3 \cdot 10^{-3}$ we can infer from eq.({\ref{shat})  bounds upon $M_\r$.  The upper bound is when $\e^2 \rightarrow \infty$.  However it is true that for lower $\e$ the large $N$ limit is valid as an approximation.  This is because the vertices in the loop expansion of the 2-point functions have a coupling of $g_{L,R}/\sqrt{L} = g^{(\prime)}/\e^2$.  For moderate $\e$ the tree-level contribution to the 2-point functions is a good approximation.  At some point the perturbative expansion breaks down as $\e$ is decreased.  In the case that $\e$ is small (and the approximations of the model are questionable) the lowest value of $M_\r$ is found.  We quote this bound for completeness.  The range of lowest bounds our model predicts is,
\eq{
M_\r \in [2.2,4.2] \textrm{TeV}
}for $\D = 2$.  An alternative way to understand this range is to say that $1/L_1 =0.9$ TeV.  Such a statement about $1/L_1$  is independent of $\e$.

To leading order in $m_{W,Z}^2/M_\r^2$  the remaining precision parameters are,
\eqs{
\hat{T}&=&0\\
\hat{U}&=&0\\
W&=&\frac{1}{4}\frac{\D^2}{\D^2-1}\e^2\hat{S}\\
X&=&\frac{1}{4}\frac{g'}{g}\frac{(2\D^2 -1)}{(\D^2-1)} \hat{S}^2 \\
Y&=&\frac{1}{4}\frac{g^2-g'^2}{g^2+g'^2}\frac{\D^2}{\D^2-1}\e^2 \hat{S}
}These expressions agree with the expectation that $X$ is sub-leading to $\hat{S}$ while $W$ and $Y$ are leading in their class of measure~\cite{LEP}.  If the $\r$ mass is taken to be TeV then the precision parameters of $\hat{S}$ and $X$ can be of order $10^{-3}$.  This fits the upper bounds.  

For the parameters $\hat{T}$ and $\hat{U}$ there is a cancellation at all orders between terms originating from the custodial symmetry of the model.  

Of most interest after $\hat{S}$  are $W$ and $Y$.  Both of these parameters  scale as $N^2$.  This means in the large-$N$ limit they both blow up.  To see the origin of this, the discussion at the end of Sec. \ref{N} is useful.  The higher order $q^2$ terms in the propagators have an extra factor of $N$ over the lower order terms.  Both $W$ and $Y$ are order 4 operators in $q^2$ and so we expect that they should blow-up at large-$N$.  The factor of $m_W^2$ in the definitions of $W$, $X$, and $Y$ give one factor of $N$ and the propagators give a second factor - so all three of these parameters should scale as $N^2$ (as found).  The parameter $X$ is also of order 4 however it can be written as $\hat{S}^2$ and so by controlling the masses at a large-$N$ but finite $N$ it does not diverge as quickly as $W$ or $Y$ which have factors of $\e^2$.  Therefore $W$, $X$ and $Y$ all blow-up at a large-$N$ but finite $N$ but the parameters of the model are adjusted in such a way as to con
 trol the spectrum which controls $X$ but not $W$ or $Y$.

A bound can  be put upon $\e$ from  $W$ so that only moderate $\mathcal{O}(1)$ values are acceptable.  As an illustration, if $\hat{S}=3\cdot 10^{-3}$ then the bound on $\e^2$ such that $W$ is no bigger than $3 \cdot 10^{-3}$ is,
\eq{
\e^2 \lesssim 4 \frac{\D^2 -1}{\D^2}.
}A consequence of choosing $\e$ within the given range is a tension with the  large-$N$ limit.  This is relevant for model building and phenomenology.   For example, the case of a `heavy $Z$' decaying into standard model $W$ bosons.  Such diagrams are difficult to compute and can be neglected in collider simulations, for example~\cite{newbosons}, by recourse to the large-$N$ limit.  In making this assumption one is predicting values of $W$ and $Y$ that are around the current upper bounds.  In order that the large-$N$ limit be a good approximation we wish to make the bound on $\e^2$ as large as possible.  This favours large $\D$.  If we consider $\e^2 \lesssim 1$ to be the point at which the description used breaks down then we have that $\D \gtrsim 2/\sqrt{3}$.  This value of $\D$ corresponds to a very large anomalous dimension, larger than previous estimates that suggest $\D =2$~\cite{wTC1, wTC3, wTC4, wTC5, wTC6, wTC7, wTC9,wTC14,wTC15,wTC16,wTC10}, these estimates put $\D$
  a
 bove the lower bound derived here. Therefore, although there are no concrete predictions for values of $\D$ which lead to walking behaviour, it is likely that suitable $\D$ exist where the large-$N$ approximation is valid in our model and the $W$ parameter satisfied.

We have shown that the model can satisfy the current bounds on precision tests.  In our set-up, after renormalisation and choosing a TeV $\r$ mass, there is only one parameter available, $\e$.  The precision tests constrain $\e$  to be of order $10^0$.   This has important consequences for model-building and the phenomenology of extra-dimension walking models.

\subsection{Comparison to an $SU(2)\times U(1)$ Model}
An alternative model to the one we study here is to gauge only the $SU(2)\times U(1)$ subgroup of the $SU(2)\times SU(2)$ global symmetry.  In this case the $\hat{T}$ and $\hat{U}$ parameters are,
\eqs{
\hat{T} &=& \frac{2}{\e^2}\frac{\D (\D-1)}{(\D+1)(2\D-1)}\left(\frac{m_Z^2}{m_W^2}-1 \right) \hat{S},\\
\hat{U}&=&\frac{1}{4}\frac{\D^3 (2+\D (2\D^3 -2 \D-1) )}{(2\D-1)(\D^2-1)^4}\left(\frac{m_Z^2}{m_W^2}-1 \right)\hat{S}^2 
}with $\hat{S}$ as defined previously. When the full $SU(2)\times SU(2)$ is gauged both of these parameters vanish.  This is the effect of the custodial symmetry.  The $\hat{T}$ parameter scales as $1$ in the large-$N$ limit and $\hat{U}$ scales as $N$.  These are the expected scalings.
\section{Comparison of the Regularised and Renormalised Cases\label{comp}}
As a demonstration of the need for holographic renormalisation, in this section an illustrative example of the model parameters is chosen to compare the holographically renormalised results with those obtained by regularising (but not renormalising) the theory.

Suppose the IR cut-off of the model is $1/L_1=0.9$ TeV and that the anomalous dimension of the condensate is such that $\D=2$.  The experimental values for the $W$ mass and couplings $g$ and $g'$ are kept fixed for this exercise.

In the holographically renormalised case, for $\e \sim \mathcal{O}(1)$, the $\r$ mass is $\mathcal{O}(3.7)$ TeV and
\eq{\label{holocase}
\hat{S} = \frac{3}{8}(m_WL_1)^2 =3\cdot 10^{-3}.
}

Now consider the computations of Sec. \ref{Divs} keeping the next-to-leading order correction in $L_0$ and without applying the counterterm $Z$ (Z=0).  In this case
\eq{
\left.\Pi_V'(q^2)\right\vert_{q^2=0} = \log \frac{L_1}{L_0}+\mathcal{O}(L_0^3),
}and for the axial case
\eq{
\left.\Pi_A'(q^2)\right\vert_{q^2=0} = -\frac{3}{16} \a^2 L_1^4+\log\frac{L_1}{L_0} +\half L_0^2 \a^2 L_1^2\log\frac{L_1}{L_0} +\mathcal{O}(L_0^3).
}After normalising the vacuum polarisations properly,
\eqs{
\hat{S} &=& \frac{g^2}{g^2+g'^2}\left(-\frac{3}{16}  \a^2 L_1^4 + \half L_0^2 \a^2 L_1^2\log\frac{L_1}{L_0} +\mathcal{O}(L_0^3)\right)\nonumber\\
&\simeq &\frac{3}{8}  m_W^2L_1^2 - L_0^2 m_W^2 \log\frac{L_1}{L_0} .
}This formula has a correction of order $L_0^2$ away from eq. (\ref{holocase}).  By choosing the UV cut-off to be very high, for example $1/L_0 = 300$ TeV, the correction to $\hat{S}$ is negligble. 

Using $1/L_0 = 300$ TeV the $\r$ mass can be computed.  The result is $M_\r =3$ TeV. 

Starting from the same action two different phenomenological scenarios have arisen depending upon whether or not holographic renormalisation is employed.  Both scenarios have the same value of $\hat{S}$ but the two spectra are different.  This picture can be understood by looking at the 4D gauge coupling.  Using the holographic renormalisation formalism it was seen, in eqs. (\ref{coupling1}) and (\ref{coupling2}), that the boundary gauge couplings are proportional to $\e g_{L,R}$.  Without the holographic renormalisation prescription the 4D coupling  is (found using eq. (\ref{norm}) with $Z=0$)
\eq{
(g^{(\prime)})^2 = \frac{1}{\log L_1/L_0}\frac{g^2_{L(R)}}{L}
}which vanishes as the UV cut-off goes to infinity -- effectively ungauging the symmetry.  Importantly the factor of $\log L_1/L_0$ is  dependent upon the precise background up to $y=L_0$ and details of the physics at the UV cut-off.  For studies on the effects of various possible  descriptions of UV physics see~\cite{delAguila:2003bh}.

In the case of this model the UV divergences originate from  integrating out the extra-dimension as the UV cut-off goes to infinity (rather than loops).  A counterterm localised on the UV boundary can be used to absorb the spurious dependence upon the UV cut-off.  This produces a renormalised theory with couplings and masses no longer dependent on the UV cut-off.  In doing so one introduces the parameter $\e$ which crucially is dependent upon the IR physics of the model.  In particular $\e$ is related to the 4D gauge coupling.  The formalism of holographic renormalisation in this context amounts to keeping a weakly gauged symmetry as  compared to the weak coupling vanishing as $1/L_0\rightarrow \infty$ and effectively rendering the symmetry global.

\section{Conclusions\label{Conc}}
Starting from a slice of AdS, we have described a prescription to obtain the propagators of a 4D theory of electroweak symmetry breaking.  The resulting model describes a strongly coupled large-$N$ gauge theory.  This has allowed us to further clarify several previously discussed points in the literature~\cite{AdSQCD1, Hirn1, Rattazzi, Grojean, Dietrich, Carone, Hirn2, Weiler,LEP,Yamawaki,Nunez,Piai}.

A bound upon the number of colors in  walking technicolor theories is commonly held to be derived from the precision parameter $\hat{S}$.  In our study the precision parameters $W$ and $Y$ are shown to blow-up in the large-$N$ limit.  Therefore the common opinion that there is an upper bound to the number of colors is confirmed in our study but for a different reason to that expected.  Usually one claims that $\hat{S}$ is proportional to $N$ and therefore there is some upper bound on $N$.  However one adjusts model parameters as $N$ increases to hold the mass spectrum fixed so the resulting effects are weak.  By comparison the $W$ and $Y$ parameters are proportional to $N\cdot  \hat{S}$ which quickly diverge as $N$ grows. 

The large-$N$ scaling of our model has been discussed in detail.    A picture that is consistent with expectation from large-$N$ QCD was found but with some additional structure originating from the introduction of the boundary terms.  In the large-$N$ limit we were able to demonstrate that the physics of the electroweak gauge bosons in  our model becomes strongly coupled, which was anticipated by previous work.

By imposing a custodial symmetry the precision parameters $\hat{T}$ and $\hat{U}$ can be made to vanish.  In our study constraints on $\hat{T}$ are removed.  

The expression for $\hat{S}$ allowed a lower bound on the $\r$ mass to be given.   One way to express this bound is to directly quote that the mass must be  greater than some value in the range  2.2  to 4.2 TeV depending upon choice of a model parameter.  An alternative is to quote the IR cut-off of the theory, which was seen to be closely related to the $\r$ mass, as being 0.9 TeV.  This result agrees with previous studies.

In clarifying the problems associated to large-$N$ within our model we were also able to demonstrate that there exists values of the anomalous dimension and appropriate $\r$ masses consistent with current estimates and experimental bounds.

\section{Acknowledgements}
The author wishes to thank G. Aarts, A. Armoni, S. Hands, G. Shore and M. Piai  for comments on the manuscript, useful discussions and frequent encouragement.  This work was supported by the STFC grant ST/F00706X/1.

\bibliography{bib}

\begin{thebibliography}{58}
\expandafter\ifx\csname natexlab\endcsname\relax\def\natexlab#1{#1}\fi
\expandafter\ifx\csname bibnamefont\endcsname\relax
  \def\bibnamefont#1{#1}\fi
\expandafter\ifx\csname bibfnamefont\endcsname\relax
  \def\bibfnamefont#1{#1}\fi
\expandafter\ifx\csname citenamefont\endcsname\relax
  \def\citenamefont#1{#1}\fi
\expandafter\ifx\csname url\endcsname\relax
  \def\url#1{\texttt{#1}}\fi
\expandafter\ifx\csname urlprefix\endcsname\relax\def\urlprefix{URL }\fi
\providecommand{\bibinfo}[2]{#2}
\providecommand{\eprint}[2][]{\url{#2}}

\bibitem[{\citenamefont{Weinberg}(1976)}]{TC1}
\bibinfo{author}{\bibfnamefont{S.}~\bibnamefont{Weinberg}},
  \bibinfo{journal}{Phys. Rev.} \textbf{\bibinfo{volume}{D13}},
  \bibinfo{pages}{974} (\bibinfo{year}{1976}).

\bibitem[{\citenamefont{Susskind}(1979)}]{TC2}
\bibinfo{author}{\bibfnamefont{L.}~\bibnamefont{Susskind}},
  \bibinfo{journal}{Phys. Rev.} \textbf{\bibinfo{volume}{D20}},
  \bibinfo{pages}{2619} (\bibinfo{year}{1979}).

\bibitem[{\citenamefont{Weinberg}(1979)}]{TC3}
\bibinfo{author}{\bibfnamefont{S.}~\bibnamefont{Weinberg}},
  \bibinfo{journal}{Phys. Rev.} \textbf{\bibinfo{volume}{D19}},
  \bibinfo{pages}{1277} (\bibinfo{year}{1979}).

\bibitem[{\citenamefont{Peskin and Takeuchi}(1990)}]{PT1}
\bibinfo{author}{\bibfnamefont{M.~E.} \bibnamefont{Peskin}} \bibnamefont{and}
  \bibinfo{author}{\bibfnamefont{T.}~\bibnamefont{Takeuchi}},
  \bibinfo{journal}{Phys. Rev. Lett.} \textbf{\bibinfo{volume}{65}},
  \bibinfo{pages}{964} (\bibinfo{year}{1990}).

\bibitem[{\citenamefont{Peskin and Takeuchi}(1992)}]{PT2}
\bibinfo{author}{\bibfnamefont{M.~E.} \bibnamefont{Peskin}} \bibnamefont{and}
  \bibinfo{author}{\bibfnamefont{T.}~\bibnamefont{Takeuchi}},
  \bibinfo{journal}{Phys. Rev.} \textbf{\bibinfo{volume}{D46}},
  \bibinfo{pages}{381} (\bibinfo{year}{1992}).

\bibitem[{\citenamefont{Holdom and Terning}(1990)}]{PT3}
\bibinfo{author}{\bibfnamefont{B.}~\bibnamefont{Holdom}} \bibnamefont{and}
  \bibinfo{author}{\bibfnamefont{J.}~\bibnamefont{Terning}},
  \bibinfo{journal}{Phys. Lett.} \textbf{\bibinfo{volume}{B247}},
  \bibinfo{pages}{88} (\bibinfo{year}{1990}).

\bibitem[{\citenamefont{Golden and Randall}(1991)}]{PT4}
\bibinfo{author}{\bibfnamefont{M.}~\bibnamefont{Golden}} \bibnamefont{and}
  \bibinfo{author}{\bibfnamefont{L.}~\bibnamefont{Randall}},
  \bibinfo{journal}{Nucl. Phys.} \textbf{\bibinfo{volume}{B361}},
  \bibinfo{pages}{3} (\bibinfo{year}{1991}).

\bibitem[{\citenamefont{Cohen and Georgi}(1989)}]{wTC1}
\bibinfo{author}{\bibfnamefont{A.}~\bibnamefont{Cohen}} \bibnamefont{and}
  \bibinfo{author}{\bibfnamefont{H.}~\bibnamefont{Georgi}},
  \bibinfo{journal}{Nuclear Physics B} \textbf{\bibinfo{volume}{314}},
  \bibinfo{pages}{7 } (\bibinfo{year}{1989}).

\bibitem[{\citenamefont{Sundrum and Hsu}(1993)}]{wTC3}
\bibinfo{author}{\bibfnamefont{R.}~\bibnamefont{Sundrum}} \bibnamefont{and}
  \bibinfo{author}{\bibfnamefont{S.~D.~H.} \bibnamefont{Hsu}},
  \bibinfo{journal}{Nucl. Phys.} \textbf{\bibinfo{volume}{B391}},
  \bibinfo{pages}{127} (\bibinfo{year}{1993}), \eprint{hep-ph/9206225}.

\bibitem[{\citenamefont{Appelquist and Sannino}(1999)}]{wTC4}
\bibinfo{author}{\bibfnamefont{T.}~\bibnamefont{Appelquist}} \bibnamefont{and}
  \bibinfo{author}{\bibfnamefont{F.}~\bibnamefont{Sannino}},
  \bibinfo{journal}{Phys. Rev.} \textbf{\bibinfo{volume}{D59}},
  \bibinfo{pages}{067702} (\bibinfo{year}{1999}), \eprint{hep-ph/9806409}.

\bibitem[{\citenamefont{Harada et~al.}(2006)\citenamefont{Harada, Kurachi, and
  Yamawaki}}]{wTC5}
\bibinfo{author}{\bibfnamefont{M.}~\bibnamefont{Harada}},
  \bibinfo{author}{\bibfnamefont{M.}~\bibnamefont{Kurachi}}, \bibnamefont{and}
  \bibinfo{author}{\bibfnamefont{K.}~\bibnamefont{Yamawaki}},
  \bibinfo{journal}{Prog. Theor. Phys.} \textbf{\bibinfo{volume}{115}},
  \bibinfo{pages}{765} (\bibinfo{year}{2006}), \eprint{hep-ph/0509193}.

\bibitem[{\citenamefont{Kurachi and Shrock}(2006)}]{wTC6}
\bibinfo{author}{\bibfnamefont{M.}~\bibnamefont{Kurachi}} \bibnamefont{and}
  \bibinfo{author}{\bibfnamefont{R.}~\bibnamefont{Shrock}},
  \bibinfo{journal}{Phys. Rev.} \textbf{\bibinfo{volume}{D74}},
  \bibinfo{pages}{056003} (\bibinfo{year}{2006}), \eprint{hep-ph/0607231}.

\bibitem[{\citenamefont{Kurachi et~al.}(2007)\citenamefont{Kurachi, Shrock, and
  Yamawaki}}]{wTC7}
\bibinfo{author}{\bibfnamefont{M.}~\bibnamefont{Kurachi}},
  \bibinfo{author}{\bibfnamefont{R.}~\bibnamefont{Shrock}}, \bibnamefont{and}
  \bibinfo{author}{\bibfnamefont{K.}~\bibnamefont{Yamawaki}},
  \bibinfo{journal}{Phys. Rev.} \textbf{\bibinfo{volume}{D76}},
  \bibinfo{pages}{035003} (\bibinfo{year}{2007}), \eprint{0704.3481}.

\bibitem[{\citenamefont{Holdom}(1985)}]{wTC8}
\bibinfo{author}{\bibfnamefont{B.}~\bibnamefont{Holdom}},
  \bibinfo{journal}{Phys. Lett.} \textbf{\bibinfo{volume}{B150}},
  \bibinfo{pages}{301} (\bibinfo{year}{1985}).

\bibitem[{\citenamefont{Yamawaki et~al.}(1986)\citenamefont{Yamawaki, Bando,
  and Matumoto}}]{wTC9}
\bibinfo{author}{\bibfnamefont{K.}~\bibnamefont{Yamawaki}},
  \bibinfo{author}{\bibfnamefont{M.}~\bibnamefont{Bando}}, \bibnamefont{and}
  \bibinfo{author}{\bibfnamefont{K.-i.} \bibnamefont{Matumoto}},
  \bibinfo{journal}{Phys. Rev. Lett.} \textbf{\bibinfo{volume}{56}},
  \bibinfo{pages}{1335} (\bibinfo{year}{1986}).

\bibitem[{\citenamefont{Bando et~al.}(1986)\citenamefont{Bando, Matumoto, and
  Yamawaki}}]{wTC10}
\bibinfo{author}{\bibfnamefont{M.}~\bibnamefont{Bando}},
  \bibinfo{author}{\bibfnamefont{K.-i.} \bibnamefont{Matumoto}},
  \bibnamefont{and} \bibinfo{author}{\bibfnamefont{K.}~\bibnamefont{Yamawaki}},
  \bibinfo{journal}{Phys. Lett.} \textbf{\bibinfo{volume}{B178}},
  \bibinfo{pages}{308} (\bibinfo{year}{1986}).

\bibitem[{\citenamefont{Appelquist et~al.}(1986)\citenamefont{Appelquist,
  Karabali, and Wijewardhana}}]{wTC13}
\bibinfo{author}{\bibfnamefont{T.~W.} \bibnamefont{Appelquist}},
  \bibinfo{author}{\bibfnamefont{D.}~\bibnamefont{Karabali}}, \bibnamefont{and}
  \bibinfo{author}{\bibfnamefont{L.~C.~R.} \bibnamefont{Wijewardhana}},
  \bibinfo{journal}{Phys. Rev. Lett.} \textbf{\bibinfo{volume}{57}},
  \bibinfo{pages}{957} (\bibinfo{year}{1986}).

\bibitem[{\citenamefont{Appelquist and Wijewardhana}(1987)}]{wTC14}
\bibinfo{author}{\bibfnamefont{T.}~\bibnamefont{Appelquist}} \bibnamefont{and}
  \bibinfo{author}{\bibfnamefont{L.~C.~R.} \bibnamefont{Wijewardhana}},
  \bibinfo{journal}{Phys. Rev.} \textbf{\bibinfo{volume}{D36}},
  \bibinfo{pages}{568} (\bibinfo{year}{1987}).

\bibitem[{\citenamefont{Yamawaki}(1996)}]{wTC15}
\bibinfo{author}{\bibfnamefont{K.}~\bibnamefont{Yamawaki}}
  (\bibinfo{year}{1996}), \eprint{hep-ph/9603293}.

\bibitem[{\citenamefont{Banks and Zaks}(1982)}]{wTC16}
\bibinfo{author}{\bibfnamefont{T.}~\bibnamefont{Banks}} \bibnamefont{and}
  \bibinfo{author}{\bibfnamefont{A.}~\bibnamefont{Zaks}},
  \bibinfo{journal}{Nucl. Phys.} \textbf{\bibinfo{volume}{B196}},
  \bibinfo{pages}{189} (\bibinfo{year}{1982}).

\bibitem[{\citenamefont{Maldacena}(1998)}]{AdSCFT1}
\bibinfo{author}{\bibfnamefont{J.~M.} \bibnamefont{Maldacena}},
  \bibinfo{journal}{Adv. Theor. Math. Phys.} \textbf{\bibinfo{volume}{2}},
  \bibinfo{pages}{231} (\bibinfo{year}{1998}), \eprint{hep-th/9711200}.

\bibitem[{\citenamefont{Gubser et~al.}(1998)\citenamefont{Gubser, Klebanov, and
  Polyakov}}]{AdSCFT2}
\bibinfo{author}{\bibfnamefont{S.~S.} \bibnamefont{Gubser}},
  \bibinfo{author}{\bibfnamefont{I.~R.} \bibnamefont{Klebanov}},
  \bibnamefont{and} \bibinfo{author}{\bibfnamefont{A.~M.}
  \bibnamefont{Polyakov}}, \bibinfo{journal}{Phys. Lett.}
  \textbf{\bibinfo{volume}{B428}}, \bibinfo{pages}{105} (\bibinfo{year}{1998}),
  \eprint{hep-th/9802109}.

\bibitem[{\citenamefont{Witten}(1998)}]{AdSCFT3}
\bibinfo{author}{\bibfnamefont{E.}~\bibnamefont{Witten}},
  \bibinfo{journal}{Adv. Theor. Math. Phys.} \textbf{\bibinfo{volume}{2}},
  \bibinfo{pages}{253} (\bibinfo{year}{1998}), \eprint{hep-th/9802150}.

\bibitem[{\citenamefont{Arkani-Hamed et~al.}(2001)\citenamefont{Arkani-Hamed,
  Porrati, and Randall}}]{AdSCFT4}
\bibinfo{author}{\bibfnamefont{N.}~\bibnamefont{Arkani-Hamed}},
  \bibinfo{author}{\bibfnamefont{M.}~\bibnamefont{Porrati}}, \bibnamefont{and}
  \bibinfo{author}{\bibfnamefont{L.}~\bibnamefont{Randall}},
  \bibinfo{journal}{JHEP} \textbf{\bibinfo{volume}{08}}, \bibinfo{pages}{017}
  (\bibinfo{year}{2001}), \eprint{hep-th/0012148}.

\bibitem[{\citenamefont{Aharony et~al.}(2000)\citenamefont{Aharony, Gubser,
  Maldacena, Ooguri, and Oz}}]{AdSCFT5}
\bibinfo{author}{\bibfnamefont{O.}~\bibnamefont{Aharony}},
  \bibinfo{author}{\bibfnamefont{S.~S.} \bibnamefont{Gubser}},
  \bibinfo{author}{\bibfnamefont{J.~M.} \bibnamefont{Maldacena}},
  \bibinfo{author}{\bibfnamefont{H.}~\bibnamefont{Ooguri}}, \bibnamefont{and}
  \bibinfo{author}{\bibfnamefont{Y.}~\bibnamefont{Oz}}, \bibinfo{journal}{Phys.
  Rept.} \textbf{\bibinfo{volume}{323}}, \bibinfo{pages}{183}
  (\bibinfo{year}{2000}), \eprint{hep-th/9905111}.

\bibitem[{\citenamefont{Da~Rold and Pomarol}(2005)}]{AdSQCD1}
\bibinfo{author}{\bibfnamefont{L.}~\bibnamefont{Da~Rold}} \bibnamefont{and}
  \bibinfo{author}{\bibfnamefont{A.}~\bibnamefont{Pomarol}},
  \bibinfo{journal}{Nucl. Phys.} \textbf{\bibinfo{volume}{B721}},
  \bibinfo{pages}{79} (\bibinfo{year}{2005}), \eprint{hep-ph/0501218}.

\bibitem[{\citenamefont{Erlich et~al.}(2005)\citenamefont{Erlich, Katz, Son,
  and Stephanov}}]{AdSQCD2}
\bibinfo{author}{\bibfnamefont{J.}~\bibnamefont{Erlich}},
  \bibinfo{author}{\bibfnamefont{E.}~\bibnamefont{Katz}},
  \bibinfo{author}{\bibfnamefont{D.~T.} \bibnamefont{Son}}, \bibnamefont{and}
  \bibinfo{author}{\bibfnamefont{M.~A.} \bibnamefont{Stephanov}},
  \bibinfo{journal}{Phys. Rev. Lett.} \textbf{\bibinfo{volume}{95}},
  \bibinfo{pages}{261602} (\bibinfo{year}{2005}), \eprint{hep-ph/0501128}.

\bibitem[{\citenamefont{Hong et~al.}(2006)\citenamefont{Hong, Yoon, and
  Strassler}}]{AdSQCD3}
\bibinfo{author}{\bibfnamefont{S.}~\bibnamefont{Hong}},
  \bibinfo{author}{\bibfnamefont{S.}~\bibnamefont{Yoon}}, \bibnamefont{and}
  \bibinfo{author}{\bibfnamefont{M.~J.} \bibnamefont{Strassler}},
  \bibinfo{journal}{JHEP} \textbf{\bibinfo{volume}{04}}, \bibinfo{pages}{003}
  (\bibinfo{year}{2006}), \eprint{hep-th/0409118}.

\bibitem[{\citenamefont{Karch et~al.}(2006)\citenamefont{Karch, Katz, Son, and
  Stephanov}}]{AdSQCD4}
\bibinfo{author}{\bibfnamefont{A.}~\bibnamefont{Karch}},
  \bibinfo{author}{\bibfnamefont{E.}~\bibnamefont{Katz}},
  \bibinfo{author}{\bibfnamefont{D.~T.} \bibnamefont{Son}}, \bibnamefont{and}
  \bibinfo{author}{\bibfnamefont{M.~A.} \bibnamefont{Stephanov}},
  \bibinfo{journal}{Phys. Rev.} \textbf{\bibinfo{volume}{D74}},
  \bibinfo{pages}{015005} (\bibinfo{year}{2006}), \eprint{hep-ph/0602229}.

\bibitem[{\citenamefont{Hong et~al.}(2005)\citenamefont{Hong, Yoon, and
  Strassler}}]{AdSQCD5}
\bibinfo{author}{\bibfnamefont{S.}~\bibnamefont{Hong}},
  \bibinfo{author}{\bibfnamefont{S.}~\bibnamefont{Yoon}}, \bibnamefont{and}
  \bibinfo{author}{\bibfnamefont{M.~J.} \bibnamefont{Strassler}}
  (\bibinfo{year}{2005}), \eprint{hep-ph/0501197}.

\bibitem[{\citenamefont{Evans et~al.}(2005)\citenamefont{Evans, Shock, and
  Waterson}}]{AdSQCD6}
\bibinfo{author}{\bibfnamefont{N.}~\bibnamefont{Evans}},
  \bibinfo{author}{\bibfnamefont{J.~P.} \bibnamefont{Shock}}, \bibnamefont{and}
  \bibinfo{author}{\bibfnamefont{T.}~\bibnamefont{Waterson}},
  \bibinfo{journal}{Phys. Lett.} \textbf{\bibinfo{volume}{B622}},
  \bibinfo{pages}{165} (\bibinfo{year}{2005}), \eprint{hep-th/0505250}.

\bibitem[{\citenamefont{Hirn and Sanz}(2005)}]{AdSQCD7}
\bibinfo{author}{\bibfnamefont{J.}~\bibnamefont{Hirn}} \bibnamefont{and}
  \bibinfo{author}{\bibfnamefont{V.}~\bibnamefont{Sanz}},
  \bibinfo{journal}{JHEP} \textbf{\bibinfo{volume}{12}}, \bibinfo{pages}{030}
  (\bibinfo{year}{2005}), \eprint{hep-ph/0507049}.

\bibitem[{\citenamefont{Katz et~al.}(2006)\citenamefont{Katz, Lewandowski, and
  Schwartz}}]{AdSQCD8}
\bibinfo{author}{\bibfnamefont{E.}~\bibnamefont{Katz}},
  \bibinfo{author}{\bibfnamefont{A.}~\bibnamefont{Lewandowski}},
  \bibnamefont{and} \bibinfo{author}{\bibfnamefont{M.~D.}
  \bibnamefont{Schwartz}}, \bibinfo{journal}{Phys. Rev.}
  \textbf{\bibinfo{volume}{D74}}, \bibinfo{pages}{086004}
  (\bibinfo{year}{2006}), \eprint{hep-ph/0510388}.

\bibitem[{\citenamefont{Brodsky and de~Teramond}(2006)}]{AdSQCD9}
\bibinfo{author}{\bibfnamefont{S.~J.} \bibnamefont{Brodsky}} \bibnamefont{and}
  \bibinfo{author}{\bibfnamefont{G.~F.} \bibnamefont{de~Teramond}},
  \bibinfo{journal}{Phys. Rev. Lett.} \textbf{\bibinfo{volume}{96}},
  \bibinfo{pages}{201601} (\bibinfo{year}{2006}), \eprint{hep-ph/0602252}.

\bibitem[{\citenamefont{Cata}(2007)}]{AdSQCD10}
\bibinfo{author}{\bibfnamefont{O.}~\bibnamefont{Cata}}, \bibinfo{journal}{Phys.
  Rev.} \textbf{\bibinfo{volume}{D75}}, \bibinfo{pages}{106004}
  (\bibinfo{year}{2007}), \eprint{hep-ph/0605251}.

\bibitem[{\citenamefont{Cs\'aki
  et~al.}(2004{\natexlab{a}})\citenamefont{Cs\'aki, Grojean, Murayama, Pilo,
  and Terning}}]{Higgsless1}
\bibinfo{author}{\bibfnamefont{C.}~\bibnamefont{Cs\'aki}},
  \bibinfo{author}{\bibfnamefont{C.}~\bibnamefont{Grojean}},
  \bibinfo{author}{\bibfnamefont{H.}~\bibnamefont{Murayama}},
  \bibinfo{author}{\bibfnamefont{L.}~\bibnamefont{Pilo}}, \bibnamefont{and}
  \bibinfo{author}{\bibfnamefont{J.}~\bibnamefont{Terning}},
  \bibinfo{journal}{Phys. Rev. D} \textbf{\bibinfo{volume}{69}},
  \bibinfo{pages}{055006} (\bibinfo{year}{2004}{\natexlab{a}}).

\bibitem[{\citenamefont{Cs\'aki
  et~al.}(2004{\natexlab{b}})\citenamefont{Cs\'aki, Grojean, Pilo, and
  Terning}}]{Higgsless2}
\bibinfo{author}{\bibfnamefont{C.}~\bibnamefont{Cs\'aki}},
  \bibinfo{author}{\bibfnamefont{C.}~\bibnamefont{Grojean}},
  \bibinfo{author}{\bibfnamefont{L.}~\bibnamefont{Pilo}}, \bibnamefont{and}
  \bibinfo{author}{\bibfnamefont{J.}~\bibnamefont{Terning}},
  \bibinfo{journal}{Phys. Rev. Lett.} \textbf{\bibinfo{volume}{92}},
  \bibinfo{pages}{101802} (\bibinfo{year}{2004}{\natexlab{b}}).

\bibitem[{\citenamefont{Cacciapaglia et~al.}(2005)\citenamefont{Cacciapaglia,
  Csaki, Grojean, and Terning}}]{Higgsless3}
\bibinfo{author}{\bibfnamefont{G.}~\bibnamefont{Cacciapaglia}},
  \bibinfo{author}{\bibfnamefont{C.}~\bibnamefont{Csaki}},
  \bibinfo{author}{\bibfnamefont{C.}~\bibnamefont{Grojean}}, \bibnamefont{and}
  \bibinfo{author}{\bibfnamefont{J.}~\bibnamefont{Terning}},
  \bibinfo{journal}{Phys. Rev.} \textbf{\bibinfo{volume}{D71}},
  \bibinfo{pages}{035015} (\bibinfo{year}{2005}), \eprint{hep-ph/0409126}.

\bibitem[{\citenamefont{Chivukula et~al.}(2002)\citenamefont{Chivukula, Dicus,
  and He}}]{Higgsless4}
\bibinfo{author}{\bibfnamefont{R.~S.} \bibnamefont{Chivukula}},
  \bibinfo{author}{\bibfnamefont{D.~A.} \bibnamefont{Dicus}}, \bibnamefont{and}
  \bibinfo{author}{\bibfnamefont{H.-J.} \bibnamefont{He}},
  \bibinfo{journal}{Phys. Lett.} \textbf{\bibinfo{volume}{B525}},
  \bibinfo{pages}{175} (\bibinfo{year}{2002}), \eprint{hep-ph/0111016}.

\bibitem[{\citenamefont{Cacciapaglia et~al.}(2004)\citenamefont{Cacciapaglia,
  Csaki, Grojean, and Terning}}]{Higgsless5}
\bibinfo{author}{\bibfnamefont{G.}~\bibnamefont{Cacciapaglia}},
  \bibinfo{author}{\bibfnamefont{C.}~\bibnamefont{Csaki}},
  \bibinfo{author}{\bibfnamefont{C.}~\bibnamefont{Grojean}}, \bibnamefont{and}
  \bibinfo{author}{\bibfnamefont{J.}~\bibnamefont{Terning}},
  \bibinfo{journal}{Phys. Rev.} \textbf{\bibinfo{volume}{D70}},
  \bibinfo{pages}{075014} (\bibinfo{year}{2004}), \eprint{hep-ph/0401160}.

\bibitem[{\citenamefont{Chivukula et~al.}(2004)\citenamefont{Chivukula,
  Simmons, He, Kurachi, and Tanabashi}}]{Higgsless6}
\bibinfo{author}{\bibfnamefont{R.~S.} \bibnamefont{Chivukula}},
  \bibinfo{author}{\bibfnamefont{E.~H.} \bibnamefont{Simmons}},
  \bibinfo{author}{\bibfnamefont{H.-J.} \bibnamefont{He}},
  \bibinfo{author}{\bibfnamefont{M.}~\bibnamefont{Kurachi}}, \bibnamefont{and}
  \bibinfo{author}{\bibfnamefont{M.}~\bibnamefont{Tanabashi}},
  \bibinfo{journal}{Phys. Rev.} \textbf{\bibinfo{volume}{D70}},
  \bibinfo{pages}{075008} (\bibinfo{year}{2004}), \eprint{hep-ph/0406077}.

\bibitem[{\citenamefont{Casalbuoni et~al.}(2007)\citenamefont{Casalbuoni,
  De~Curtis, Dominici, and Dolce}}]{Higgsless7}
\bibinfo{author}{\bibfnamefont{R.}~\bibnamefont{Casalbuoni}},
  \bibinfo{author}{\bibfnamefont{S.}~\bibnamefont{De~Curtis}},
  \bibinfo{author}{\bibfnamefont{D.}~\bibnamefont{Dominici}}, \bibnamefont{and}
  \bibinfo{author}{\bibfnamefont{D.}~\bibnamefont{Dolce}},
  \bibinfo{journal}{JHEP} \textbf{\bibinfo{volume}{08}}, \bibinfo{pages}{053}
  (\bibinfo{year}{2007}), \eprint{0705.2510}.

\bibitem[{\citenamefont{Nunez et~al.}(2008)\citenamefont{Nunez, Papadimitriou,
  and Piai}}]{Nunez}
\bibinfo{author}{\bibfnamefont{C.}~\bibnamefont{Nunez}},
  \bibinfo{author}{\bibfnamefont{I.}~\bibnamefont{Papadimitriou}},
  \bibnamefont{and} \bibinfo{author}{\bibfnamefont{M.}~\bibnamefont{Piai}}
  (\bibinfo{year}{2008}), \eprint{0812.3655}.

\bibitem[{\citenamefont{Hirn and Sanz}(2007)}]{Hirn1}
\bibinfo{author}{\bibfnamefont{J.}~\bibnamefont{Hirn}} \bibnamefont{and}
  \bibinfo{author}{\bibfnamefont{V.}~\bibnamefont{Sanz}},
  \bibinfo{journal}{JHEP} \textbf{\bibinfo{volume}{03}}, \bibinfo{pages}{100}
  (\bibinfo{year}{2007}), \eprint{hep-ph/0612239}.

\bibitem[{\citenamefont{Piai}(2006)}]{Piai}
\bibinfo{author}{\bibfnamefont{M.}~\bibnamefont{Piai}} (\bibinfo{year}{2006}),
  \eprint{hep-ph/0608241}.

\bibitem[{\citenamefont{Barbieri
  et~al.}(2004{\natexlab{a}})\citenamefont{Barbieri, Pomarol, and
  Rattazzi}}]{Rattazzi}
\bibinfo{author}{\bibfnamefont{R.}~\bibnamefont{Barbieri}},
  \bibinfo{author}{\bibfnamefont{A.}~\bibnamefont{Pomarol}}, \bibnamefont{and}
  \bibinfo{author}{\bibfnamefont{R.}~\bibnamefont{Rattazzi}},
  \bibinfo{journal}{Phys. Lett.} \textbf{\bibinfo{volume}{B591}},
  \bibinfo{pages}{141} (\bibinfo{year}{2004}{\natexlab{a}}),
  \eprint{hep-ph/0310285}.

\bibitem[{\citenamefont{Agashe et~al.}(2007)\citenamefont{Agashe, Csaki,
  Grojean, and Reece}}]{Grojean}
\bibinfo{author}{\bibfnamefont{K.}~\bibnamefont{Agashe}},
  \bibinfo{author}{\bibfnamefont{C.}~\bibnamefont{Csaki}},
  \bibinfo{author}{\bibfnamefont{C.}~\bibnamefont{Grojean}}, \bibnamefont{and}
  \bibinfo{author}{\bibfnamefont{M.}~\bibnamefont{Reece}},
  \bibinfo{journal}{JHEP} \textbf{\bibinfo{volume}{12}}, \bibinfo{pages}{003}
  (\bibinfo{year}{2007}), \eprint{0704.1821}.

\bibitem[{\citenamefont{Dietrich et~al.}(2005)\citenamefont{Dietrich, Sannino,
  and Tuominen}}]{Dietrich}
\bibinfo{author}{\bibfnamefont{D.~D.} \bibnamefont{Dietrich}},
  \bibinfo{author}{\bibfnamefont{F.}~\bibnamefont{Sannino}}, \bibnamefont{and}
  \bibinfo{author}{\bibfnamefont{K.}~\bibnamefont{Tuominen}},
  \bibinfo{journal}{Phys. Rev. D} \textbf{\bibinfo{volume}{72}},
  \bibinfo{pages}{055001} (\bibinfo{year}{2005}).

\bibitem[{\citenamefont{Carone et~al.}(2007)\citenamefont{Carone, Erlich, and
  Tan}}]{Carone}
\bibinfo{author}{\bibfnamefont{C.~D.} \bibnamefont{Carone}},
  \bibinfo{author}{\bibfnamefont{J.}~\bibnamefont{Erlich}}, \bibnamefont{and}
  \bibinfo{author}{\bibfnamefont{J.~A.} \bibnamefont{Tan}},
  \bibinfo{journal}{Phys. Rev.} \textbf{\bibinfo{volume}{D75}},
  \bibinfo{pages}{075005} (\bibinfo{year}{2007}), \eprint{hep-ph/0612242}.

\bibitem[{\citenamefont{Hirn and Sanz}(2006)}]{Hirn2}
\bibinfo{author}{\bibfnamefont{J.}~\bibnamefont{Hirn}} \bibnamefont{and}
  \bibinfo{author}{\bibfnamefont{V.}~\bibnamefont{Sanz}},
  \bibinfo{journal}{Phys. Rev. Lett.} \textbf{\bibinfo{volume}{97}},
  \bibinfo{pages}{121803} (\bibinfo{year}{2006}), \eprint{hep-ph/0606086}.

\bibitem[{\citenamefont{Cacciapaglia et~al.}(2008)}]{Weiler}
\bibinfo{author}{\bibfnamefont{G.}~\bibnamefont{Cacciapaglia}}
  \bibnamefont{et~al.}, \bibinfo{journal}{JHEP} \textbf{\bibinfo{volume}{04}},
  \bibinfo{pages}{006} (\bibinfo{year}{2008}), \eprint{0709.1714}.

\bibitem[{\citenamefont{Haba et~al.}(2008)\citenamefont{Haba, Matsuzaki, and
  Yamawaki}}]{Yamawaki}
\bibinfo{author}{\bibfnamefont{K.}~\bibnamefont{Haba}},
  \bibinfo{author}{\bibfnamefont{S.}~\bibnamefont{Matsuzaki}},
  \bibnamefont{and} \bibinfo{author}{\bibfnamefont{K.}~\bibnamefont{Yamawaki}},
  \bibinfo{journal}{Prog. Theor. Phys.} \textbf{\bibinfo{volume}{120}},
  \bibinfo{pages}{691} (\bibinfo{year}{2008}), \eprint{0804.3668}.

\bibitem[{\citenamefont{Barbieri
  et~al.}(2004{\natexlab{b}})\citenamefont{Barbieri, Pomarol, Rattazzi, and
  Strumia}}]{LEP}
\bibinfo{author}{\bibfnamefont{R.}~\bibnamefont{Barbieri}},
  \bibinfo{author}{\bibfnamefont{A.}~\bibnamefont{Pomarol}},
  \bibinfo{author}{\bibfnamefont{R.}~\bibnamefont{Rattazzi}}, \bibnamefont{and}
  \bibinfo{author}{\bibfnamefont{A.}~\bibnamefont{Strumia}},
  \bibinfo{journal}{Nucl. Phys.} \textbf{\bibinfo{volume}{B703}},
  \bibinfo{pages}{127} (\bibinfo{year}{2004}{\natexlab{b}}),
  \eprint{hep-ph/0405040}.

\bibitem[{\citenamefont{Skenderis}(2002)}]{Skenderis}
\bibinfo{author}{\bibfnamefont{K.}~\bibnamefont{Skenderis}},
  \bibinfo{journal}{Class. Quant. Grav.} \textbf{\bibinfo{volume}{19}},
  \bibinfo{pages}{5849} (\bibinfo{year}{2002}), \eprint{hep-th/0209067}.

\bibitem[{\citenamefont{Da~Rold and Pomarol}(2006)}]{DaRold:2005vr}
\bibinfo{author}{\bibfnamefont{L.}~\bibnamefont{Da~Rold}} \bibnamefont{and}
  \bibinfo{author}{\bibfnamefont{A.}~\bibnamefont{Pomarol}},
  \bibinfo{journal}{JHEP} \textbf{\bibinfo{volume}{01}}, \bibinfo{pages}{157}
  (\bibinfo{year}{2006}), \eprint{hep-ph/0510268}.

\bibitem[{\citenamefont{Hong and Yee}(2006)}]{Hong}
\bibinfo{author}{\bibfnamefont{D.~K.} \bibnamefont{Hong}} \bibnamefont{and}
  \bibinfo{author}{\bibfnamefont{H.-U.} \bibnamefont{Yee}},
  \bibinfo{journal}{Phys. Rev.} \textbf{\bibinfo{volume}{D74}},
  \bibinfo{pages}{015011} (\bibinfo{year}{2006}), \eprint{hep-ph/0602177}.

\bibitem[{\citenamefont{Piai and Round}(2009)}]{newbosons}
\bibinfo{author}{\bibfnamefont{M.}~\bibnamefont{Piai}} \bibnamefont{and}
  \bibinfo{author}{\bibfnamefont{M.}~\bibnamefont{Round}}
  (\bibinfo{year}{2009}), \eprint{0904.1524}.

\bibitem[{\citenamefont{del Aguila et~al.}(2003)\citenamefont{del Aguila,
  Perez-Victoria, and Santiago}}]{delAguila:2003bh}
\bibinfo{author}{\bibfnamefont{F.}~\bibnamefont{del Aguila}},
  \bibinfo{author}{\bibfnamefont{M.}~\bibnamefont{Perez-Victoria}},
  \bibnamefont{and} \bibinfo{author}{\bibfnamefont{J.}~\bibnamefont{Santiago}},
  \bibinfo{journal}{JHEP} \textbf{\bibinfo{volume}{02}}, \bibinfo{pages}{051}
  (\bibinfo{year}{2003}), \eprint{hep-th/0302023}.

\end{thebibliography}
\appendix
\section{$SU(2)\times SU(2) \rightarrow SU(2)\times U(1)$ Breaking\label{appendix}}
It is useful to show explicitly that eq. (\ref{CTaction}) can be written in a manifestly gauge invariant form.  For this exercise it is best to work in the $(L,R)$ basis.

Working from the action eq.~(\ref{action}) and in the unitary gauge, one obtains the following UV boundary terms
\eq{
-\int d^4q  \half q^2\Tr \left[\frac{L}{y} \left(L_\m \partial_y L_\n + R_\m \partial_y R_\n\right)P^{\m\n}\right]_{y=L_0}.
}To renormalise the divergences in these terms add a localised kinetic (counter-)term for the $L_\m$ field and a triplet of real scalar fields, $H=H^a \s^a/2$, that transforms under the adjoint representation of $SU(2)_R$,
\eqs{
S_{C.T.} &=& \iint \sqrt{g}dyd^4x \d(y-L_0)\,\bigg[Z g^{\m\r}g^{\n\s}\bigg(- \half \Tr L_{\m\n}L_{\r\s}  \bigg) + g^{\m\n}\Tr (D_\m H D_\n H)\nonumber\\&-& \frac{Z}{4v_H^2}g^{\m\r}g^{\n\s} \Tr (R_{\m\n}H)\Tr(R_{\r\s}H )
+\l \left(\Tr HH-\half v_H^2\right)^2\bigg],\\
D_\m H  &=& \partial_\m H +ig_R ( R_\m H - HR_\m),
}Notice that $S_{C.T.}$, as defined above, and eq. (\ref{action}) are both written in a manifestly gauge invariant way.   

Give $H$ a VEV in the 3-direction, 
\eq{
\langle H \rangle = \frac{1}{2}v_H \begin{pmatrix}1&0\\0&-1\end{pmatrix}
}for $v_H$ the VEV size parameter and take the $\l \rightarrow \infty$ limit.  In doing so the scalar field $H$ decouples.  This is the limit considered throughout this work.  The kinetic term for $H$  gives a mass to the $R^{1,2}$ fields proportional to $v_H g_R$ while $R^3$ remains massless.  We take $v_H$ small in this work so that the effects of $v_H$ are negligble.

Substitute the VEV into the counterterm and concentrate on
\eq{
-\frac{Z}{4v_H^2}g^{\m\r}g^{\n\s} \Tr (R_{\m\n}H)\Tr(R_{\r\s}H ) = -\frac{Z}{4}g^{\m\r}g^{\n\s} R_{\m\n}^3R^3_{\r\s},
}which matches the $R$-field term in eq. (\ref{CTaction}).

\end{document}